\documentclass{basi}
\usepackage{txfonts}
\usepackage{graphicx}

\def\bea{\begin{eqnarray}}
\def\eea{\end{eqnarray}}
\def \no {\nonumber}
\def \A {{\cal A}}
\def \n {{\bf{\hat n}}}
\def  \R {{\bf R}}
\def \Om {\Omega}
\def \X {{\bf{\hat X}}}
\def \Y {{\bf {\hat Y}}}

\begin{document}
\title[Gravitational wave astronomy]{Gravitational wave astronomy - astronomy of the 21$^{\rm st}$ century}
\author[S.~V.~Dhurandhar]%
       {S.~V.~Dhurandhar \thanks{e-mail:sanjeev@iucaa.ernet.in} \\
       Inter University Centre for Astronomy \& Astrophysics, Ganeshkhind, Pune - 411 007.}

\pubyear{2011}
\volume{39}
\pagerange{\pageref{firstpage}--\pageref{lastpage}}
\date{Received 2011 February 11; accepted 2011 February 27}

\maketitle
\label{firstpage}

\begin{abstract}
An enigmatic prediction of Einstein's general theory of relativity is gravitational waves. 
With the observed decay in the orbit of the Hulse-Taylor binary pulsar agreeing within a 
fraction of a percent with the theoretically computed decay from Einstein's theory, the 
existence of gravitational waves was firmly established. Currently there is a worldwide 
effort to detect gravitational waves with interferometric gravitational wave observatories 
or detectors and several such detectors have been built or being built. The initial detectors 
have reached their design sensitivities and now the effort is on to construct advanced detectors 
which are expected to detect gravitational waves from astrophysical sources. The era of  
gravitational wave astronomy has arrived. This article describes the worldwide effort which 
includes the  effort on the Indian front - the IndIGO project -, the principle underlying 
interferometric detectors both on ground and in space, the principal noise sources that plague 
such detectors, the astrophysical sources of gravitational waves that one expects to detect by 
these detectors and some glimpse of the data analysis methods involved in extracting the very 
weak gravitational wave signals from detector noise. 

\end{abstract}


\section{Introduction}\label{s:intro}

In the past half a century or so, astronomy has been revolutionised by several unexpected discoveries 
because of the plethora of windows being opened in various bands of the electromagnetic spectrum. 
To name a few, the cosmic microwave background and the discovery of pulsars in radio band, gamma-ray 
bursts, X-ray objects, all go to show that whenever a new window has been opened, startling discoveries 
have followed. Yet another window to the Universe should soon open up in few years time - the 
gravitational wave (GW) window. Not only will this window test Einstein's general theory of relativity, 
but also provide direct evidence for black holes, and more generally test general relativity in the strong 
field regime. Just as the other windows to the Universe have brought in unexpected discoveries, it is not 
unreasonable to expect the same in this case also - and more so, because this involves changing the 
physical interaction from electromagnetic to gravitational.  

The existence of gravitational waves predicted by the theory of 
general relativity, has long been verified `indirectly' through the 
observations of Hulse and Taylor (Hulse \& Taylor 1975; Taylor 1994). The inspiral of 
the members of the
binary pulsar system named after them has been successfully accounted for in 
terms of the back-reaction due to the radiated gravitational waves - the observational results and the 
theory agree with each other within a fraction of a percent.
However, detecting such waves directly with the help of detectors based either on  
ground or in space has not been possible so far. 

The key to gravitational wave detection is the very precise measurement of
small changes in distance. For laser interferometers, this is the distance
between pairs of mirrors hanging at either end of two long, mutually
perpendicular vacuum chambers. A GW passing through the
instrument will shorten one arm while lengthening the other. By using an
interferometric design, the relative change in length of the two arms can be
measured, thus signalling the passage of a GW at the
detector site. GW detectors produce an enormous volume of 
output consisting mainly of noise from a host of sources both environmental and intrinsic. 
Buried in this noise will be the GW signature. Sophisticated data analysis
techniques are needed to optimally extract the GW signal from the interferometric data. 
IUCAA has made significant contributions in this area.

\section{Interferometric detection of GW}

Historically with pioneering efforts of Weber in the 1960s, the detectors were resonant 
bar detectors which were suspended, seismically isolated, aluminium cylinders. The later 
versions were cooled to extremely low temperatures - ultracryogenic - to suppress the thermal 
noise. There are also spherical resonant mass  detectors being constructed/operating. However, 
these ideas although interesting and useful in their niche, have their limitations. In this 
article we will confine ourselves to the interferometric detectors.  

\subsection{The principle of interferometric detection}

A weak GW is described by a metric perturbation 
$h_{\mu \nu}$ in general relativity. Typically, for the astrophysical GW 
sources which are amenable to detection, $h_{\mu \nu} \sim 10^{-22}$. In the transverse-traceless  
gauge, the $h_{\mu \nu}$ can be expressed in terms of just two amplitudes, 
$h_+$ and $h_{\times}$, called the `plus' and `cross' polarisations. 
If a weak monochromatic gravitational wave of + polarisation is incident 
on a ring of test-particles, the ring is deformed into an ellipse as shown at the top of 
Figure 1. Phases, a quarter cycle apart, of the GW are shown in the Figure. For 
the $\times$ polarisation the ellipses are rotated by an angle of $45^\circ$. A general wave 
is a linear combination of the two polarisations.

\begin{figure}[tbh!]
\centerline{\includegraphics[width=6.0cm]{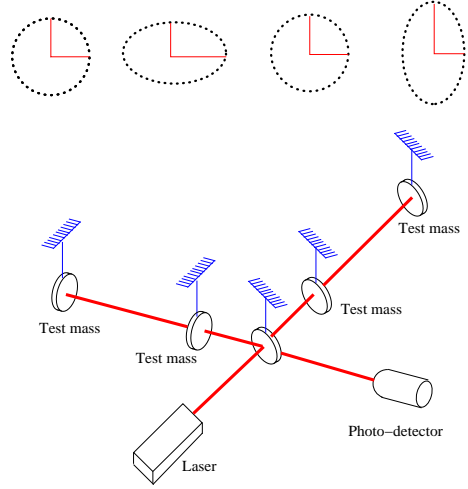}}
\caption{Upper: A circular ring of test particles is deformed into an ellipse by an incident GW. 
Phases, a quarter of a cycle apart are shown for the + polarisation. The length change in 
the interferometric arms is also shown schematically. Lower: a schematic diagram of an interferometer is drawn.}
\label{fig1}
\end{figure}

At the bottom of Figure 1, a schematic of the interferometer is depicted. If the change in  the 
armlength $L$ is $\delta L$, then, 
\begin{equation}
\delta L \sim h L,
\end{equation}
where $h$ is a typical component of the metric perturbation.  

For a GW source, $h$ can be estimated from the well-known Landau-Lifschitz 
quadrupole formula. The GW amplitude $h$ is related to the second time derivative 
of the quadrupole moment (which has dimensions of energy) of the source:  
\begin{equation}
 h \sim {4\over r}~ {G\over c^4} ~ E^{\rm kinetic}_{\rm nonspherical},  
\end{equation} 
where $r$ is the distance to the source, $G$ is the gravitational constant, 
$c$ is the speed of light and $E^{\rm kinetic}_{\rm nonspherical}$ is the 
kinetic energy in the {\it nonspherical} motion of the source.
If we consider $E^{\rm kinetic}_{\rm nonspherical} / c^2$ a fraction   
of a solar mass and the distance to the source ranging from galactic 
scale of tens of kpc to cosmological distances of Gpc, then $h$ ranges from 
$10^{-17}$ to $10^{-22}$. These numbers then set the scale for the sensitivities 
at which the detectors must operate. The factor of 4 is also important given the 
weakness of the interaction and the subsequent signal extraction from detector noise.

\subsection{Ground-based interferometric detectors}

\begin{figure}[tbh!]
\centerline{\includegraphics[width=12cm]{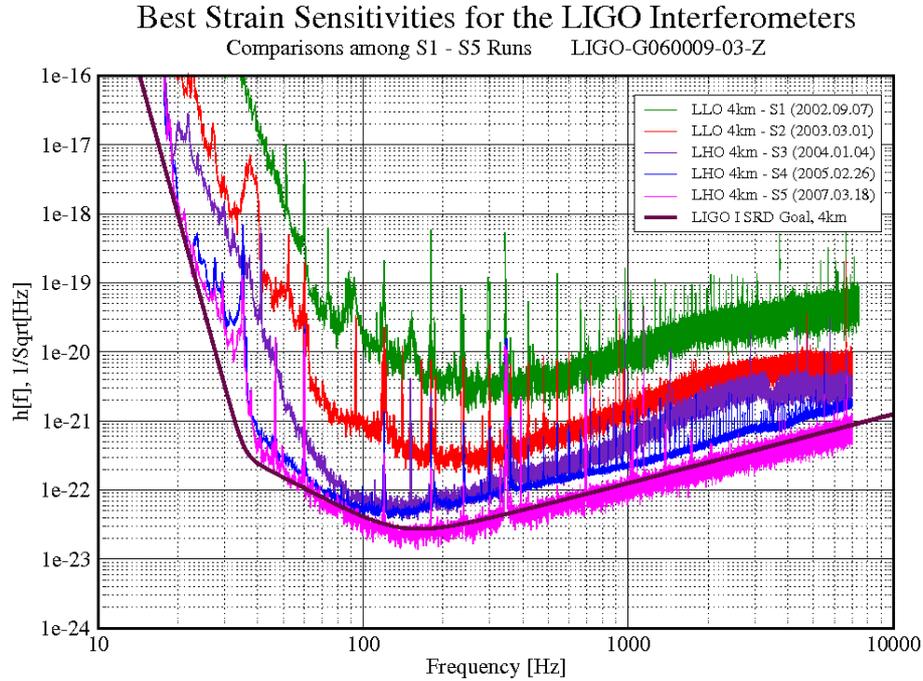}}
\caption{The figure shows the sensitivity achieved by LIGO detectors by March 2007. This sensitivity level 
has been surpassed in later operations (Gonzalez 2005; see LIGO website for the S5 curve).}
\label{fig2}
\end{figure}

There are a host of noise sources in ground-based interferometric detectors which contaminate 
the data. At low frequencies there is the seismic noise. The seismic isolation is 
a sequence of stages consisting of springs/pendulums and heavy masses. Each 
stage has a low resonant frequency about a fraction of a Hz. The seismic 
isolation acts as a low pass filter, attenuating high frequencies, but low frequencies 
get through. This results in a `noise wall' at low frequencies and marks the lower end of 
the detector bandwidth. It is about 40 Hz for initial detectors but will go down to 
10 Hz for advanced detectors increasing the bandwidth. At mid-frequencies up to a few hundred Hz, 
the thermal noise is important and is due to the thermal excitations 
both in the test masses - the mirrors - as well as the seismic suspensions. Currently, this 
seems to be the noise hardest to suppress. The natural modes of the mirrors and the suspension 
are driven by the thermal excitations. One `solution' is to cool the mirrors/suspensions, but 
this has its own problems. Nevertheless, the Japanese have planned a detector doing just 
this - the Large-scale Cryogenic Gravitational-wave Telescope (LCGT) which has been funded recently.  
At high frequencies the shot noise from the laser dominates. This noise is due 
to the quantum nature of light. From photon counting statistics and the uncertainty 
principle, the phase fluctuation is inversely proportional to the square root of 
the mean number of photons arriving during a period of the wave. So increasing the laser power 
and hence the mean number of photons during a given period of the wave tends to reduce this noise. 
Apart from these main noise sources there are other noise sources, an important one among them is 
gravity gradient noise which cannot be screened and occurs only at low frequencies. The slowly changing 
gravity gradients are due to natural causes (such as clouds moving in the sky, changes in atmospheric density) 
or are manmade.  Thus long arm lengths, high laser power, and extremely well-controlled laser stability 
are essential to reach the requisite sensitivity. Figure~\ref{fig2} shows the sensitivity achieved 
by the initial LIGO detectors (Gonzalez 2005) when the actual noise in the detectors reached 
theoretical design sensitivity (shown by the bold curve). The sensitivity has continued to improve with time. 

\subsection{The worldwide network of ground-based interferometric gravitational wave observatories}

The USA has been at the forefront in building large scale detectors. The LIGO project (Abramovici et al. 1992) 
has built three detectors, two of armlength 4 km and one of armlength 2 km at two sites about 3000 km 
apart at Hanford, Washington and at Livingston, Louisiana. The 2 km detector is at Hanford. 
These initial detectors have had several science runs and the design sensitivity has not only been 
reached but surpassed. The goal of this initial stage was mainly to vindicate the technologies involved in 
attaining the design sensitivities. Now the next phase is to build advanced detectors with state of the art 
technologies which will be capable of observing GW sources and doing GW astronomy. With these future goals 
a radical decision has been taken by the LIGO project, that of building one of its detectors in 
Australia - that is LIGO will build two advanced detectors in US and partially fund a full scale detector in 
Australia with advanced design. This detector is called LIGO-Australia and will be built in collaboration 
with the Australians who already have an interferometric facility at Gingin near Perth - the 
AIGO (Australian Interferometric Gravitationalwave Observatory) project. The reason for this decision 
by the US is clear - it is to increase the baseline and have a detector far removed from other detectors 
on Earth, which has several advantages, such as improving the localization of the GW source.

In Europe the large-scale project is the VIRGO project (Bradaschia et al. 1990) of Italy and France 
which has built a 3 km armlength detector. After commissioning of the project in 2007, it also had science 
runs. The GEO600 (Danzmann et al. 1995) is a German-British project and whose detector has been built 
near Hannover, Germany with an armlength of 600 metres. One of the goals of GEO600 is to develop advanced 
technologies required for the next generation detectors with the aim of achieving higher sensitivity. 

Japan was the first (around 2000) to have a large scale detector of 300 m armlength - the TAMA300 detector 
under the TAMA project (Tsubono 1995) - operating continuously at high sensitivity in the range of 
$h \sim 10^{-20}$. Now Japan plans to construct a cryogenic inteferometric detector called the LCGT 
(Large-scale Cryogenic Gravitational wave Telescope; Kuroda 2006) which has been recently funded. The 
purpose of the cryogenics is to quell the thermal noise. But this technnology is by no means straight 
forward and will test the skills of the experimenters.

Australia is looking for international partners, because of LIGO-Australia. Given the twenty year 
old legacy in GW data analysis at IUCAA, Pune and waveform modelling at RRI, Bangalore, Australia would 
welcome the Indians as partners in this endeavour. Recently, about two years ago, an Indian Initiative 
in Gravitational Wave Astronomy (IndIGO) has begun whose goal is to promote and foster gravitational wave 
astronomy in India and join in the worldwide quest to observe gravitational waves. Apart from the data 
analysis this initiative includes the all important experimental aspect. Accordingly a modest beginning 
has been made by IndIGO with TIFR, Mumbai approving a 3 metre prototype on which Indian experimenters 
can get first hand experience and develop expert manpower. This project has already been funded. Concurrently, 
an MOU with Australia has been signed which purports to ask for funding from Indian agencies in parallel 
with Australia. An IndIGO consortium has been formed with scientists from leading institutions such as TIFR, 
RRCAT, RRI, IUCAA, IISERs, Delhi University and CMI, and also including scientists (mainly Indian) working 
abroad. The current strength of the consortium is about 25 scientists. In order to further this effort the 
first goal is to muster up sufficient expert and skilled manpower which will launch this activity. It will 
mean India getting into this worldwide challenging experiment.

Besides the current projects, studies have begun for third generation detectors which will include further 
advanced technologies to enhance the sensitivities of GW detectors to reach out farther in the sky; 
the Einstein Telescope (ET) is just such a future goal. 

\subsection{Space-based detectors: the LISA project}

A  natural limit occurs on  decreasing the lower frequency cut-off beyond $\sim$10 Hz  
because it is not practical to  increase the arm-lengths on ground and also  because 
of the gravity gradient noise  
which is difficult to eliminate below $10$ Hz. Thus, the ground based 
interferometers  will not be sensitive below 
the limiting frequency of $\sim$10~Hz.  But on the other hand,  there exist 
in the cosmos, interesting astrophysical GW sources which emit GW below this
frequency such as the galactic binaries, massive and super-massive black hole 
binaries. If we wish to observe these sources, we need to go to lower 
frequencies. The solution is to build an interferometer in space, where such 
noises will be absent and  allow the detection of GW in the low frequency 
regime.  {\bf LISA } -~{\it Laser Interferometric Space  Antenna }~- is a 
proposed ESA-NASA mission which will use coherent laser beams exchanged between three
identical spacecraft forming a giant~(almost) equilateral triangle of side
5$\times$10$^6$ kilometers to observe and detect low-frequency cosmic 
GW. \footnote{             
http://sci.esa.int/science-e/www/area/index.cfm?fareaid=27; ~~~http://lisa.gsfc.nasa.gov }
The ground-based detectors and LISA complement each other in the  observation 
of GW in an essential way, analogous to the way optical, radio, X-ray,  
$\gamma$-ray observations do for electromagnetic waves. As these 
detectors begin  to operate, a new era of {\it gravitational astronomy} is 
on the horizon and a radically  different view of the Universe is expected 
to be revealed. There are also further space projects being considered.

\begin{figure}
\centerline{\includegraphics[width=8cm,angle=90]{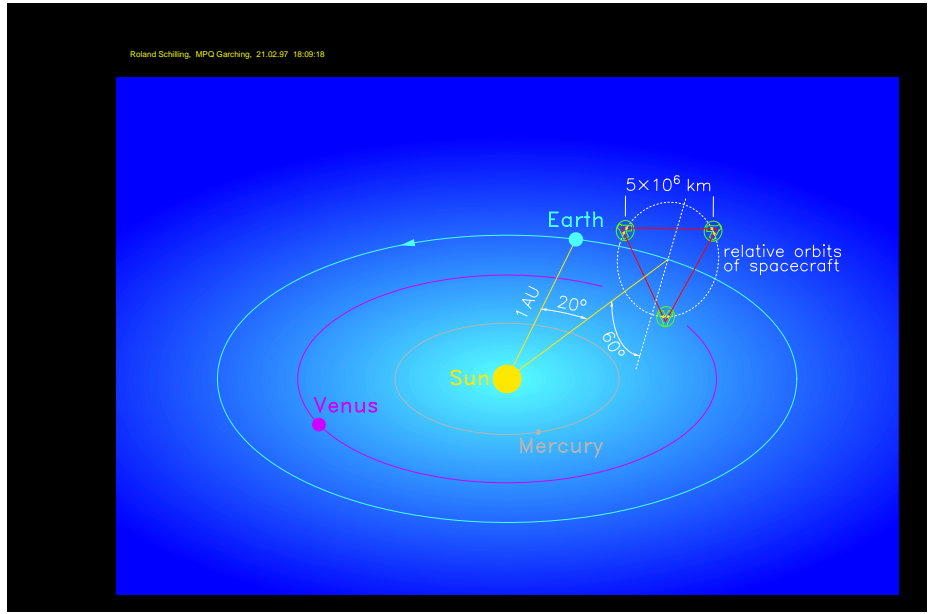}}
\caption{LISA orbital configuration around the Sun, describing a cone with
$60^{\circ}$ half opening angle. The centroid of the triangle follows
an Earth-like orbit trailing 20 degrees behind (Bender et al. 1998).  }
\label{LISA}
\end{figure}

LISA consists of three spacecrafts, flying five million kilometres apart, in 
an equilateral triangle. The spacecrafts are maintained drag-free by a complex
system of accelerometers and micro-propellers. Each spacecraft will 
carry two optical assemblies that contain the main optics and a free-falling 
inertial sensor.  The light sent out by a laser in one spacecraft is
received by the telescope on the distant spacecraft. The incoming light 
from the distant spacecraft is then mixed with the in-house laser and the 
differential phase is recorded. This defines one elementary data stream. 
There are thus six elementary data streams which are formed by going 
clockwise and anti-clockwise around the LISA triangle. Suitable combinations 
of these elementary data streams can be used to optimally extract the GW 
signal from the instrumental noise. In other words, LISA is basically a 
giant Michelson interferometer placed  in space, with a third arm added 
to give independent information on the two  gravitational wave polarisations, 
and for redundancy. The distance between the spacecrafts - the interferometer 
arm-length - determines the frequency  range in which LISA can make 
observations; it was carefully chosen to allow for the observations of most 
of the interesting sources of gravitational radiation.  Each spacecraft 
revolves  in its own heliocentric orbit. The centre of LISA's triangle will 
follow Earth's orbit around the Sun, trailing 20 degrees behind. It will 
maintain a distance of 1 AU~(astronomical unit) from the Sun, the average 
distance between the Earth and the Sun (Figure~\ref{LISA}).
The spacecrafts rotate in a circle drawn through the vertices of the 
triangle and the LISA constellation as a whole revolves around the Sun.
LISA's operational position was chosen as a compromise between the
need to minimise the effects on the spacecrafts of changes in the Earth's
gravitational field and the need to be close enough to the Earth for
easy communication. 

LISA will observe low-frequency GW in the range 0.1 mHz to 0.1 Hz. Since astrophysical systems are 
generally large and in spite of high velocities do not change their quadrupole moment too quickly, 
the Universe is richly populated with sources in this frequency band. Also the masses that produce 
GW in this frequency band are generally large and thus produce stronger GW  than those in 
ground-based detectors, leading to high signal-to-noise ratios (SNR). The signals for LISA arise 
from a large variety of phenomena, such as merging massive and supermassive black holes, vibrating black holes 
(quasi-normal modes), stellar mass objects falling into massive and supermassive black holes and GWs of 
cosmological origin. The high SNRs of these signals imply detailed and accurate information which can 
test general relativity and its ramifications to unprecedented accuracies. Astrophysics of various 
objects like compact binaries, stellar remnants can be studied and LISA observations can provide 
useful clues to events in the early Universe (Bender \& Hils 1997; Nelemans, 
Yungelson \& Portegies Zwart 2001; Postnov \& Prokhorov 1998; Hills \& Bender 2000).

LISA sensitivity is limited by several noise sources. A major noise source 
is the    laser phase~(frequency) noise which arises due to phase fluctuations 
of the master laser.   Amongst the important noise sources, laser phase noise 
is expected to be several orders of magnitude larger than other noises in the 
instrument. The current   stabilisation schemes estimate this noise to 
about $\Delta \nu /\nu_0 \simeq 3 \times 10^{-14}/\sqrt{\rm Hz}$, where 
$\nu_0$ is the  frequency of the laser and $\Delta \nu$ the fluctuation in 
frequency. If the laser frequency noise can be suppressed then the noise 
floor is determined by the   optical-path noise which acts like fluctuations 
in the lengths of optical paths and the residual acceleration of proof masses 
resulting from imperfect  shielding of the drag-free system. The noise floor 
is then at an effective GW strain  sensitivity $h \sim 10^{-21}$ or $10^{-22}$. 
Thus, cancelling the laser frequency noise is vital if LISA is to reach the 
requisite  sensitivity.

In ground-based detectors the arms are chosen to be of equal length so that
the laser light experiences identical delay in each arm of the interferometer.
This arrangement precisely cancels the laser frequency/phase noise at the
photodetector.  However, in LISA it is impossible to achieve equal distances between 
spacecrafts and also the data are taken at a phasemeter as a beat note between the 
local oscillator and the incoming beam coming from a spacecraft 5 million km away. 
In LISA, six data streams arise from the exchange of laser beams between the 
three spacecrafts - it is not possible to bounce laser beams between different spacecrafts, 
as is done in ground-based detectors. The technique of time-delay interferometry (TDI) 
is used (Armstrong, Estabrook \& Tinto 1999; Estabrook, Tinto \& Armstrong 2000) which 
combines the recorded data with suitable time-delays 
corresponding to the three arm-lengths of the giant triangular interferometer. An original approach to 
this problem was taken by IUCAA. A {\it systematic method} based on modules over polynomial rings 
has been successfully formulated which is most appropriate for this problem 
(Dhurandhar, Nayak \& Vinet 2000; 2010).
The method uses the redundancy in the data to suppress the laser frequency noise.  
 
\section{General discussion of GW sources}

\subsection{GW sources}
 
Several types of GW sources have been envisaged which could be directly 
observed by Earth-based detectors: 
(i) Burst sources -- such as binary systems consisting of neutron stars and/or 
black holes in their inspiral phase or merger phase;  
supernova explosions -- whose signals last for a time 
between a few milli-seconds and a few minutes, 
much shorter, than
the typical observation time; (ii) stochastic backgrounds of radiation,
of either primordial or astrophysical origin, and 
(iii) continuous wave  sources -- e.g. rapidly rotating non-axisymmetric 
neutron stars -- where a weak sinusoidal signal is continuously emitted. 
As one sees from the discussion that follows, the strengths of these sources are usually 
well below or even way below, the mean noise level in the detectors either currently 
operating or even for those planned in the near future - the advanced detectors. This situation 
makes the expert data analysis all the more vital, firstly in detecting the source, and secondly 
and more importantly in extracting astrophysical information about it. 

Inspiraling binaries have been considered highly promising sources not only because of 
the enormous GW energy they emit, but also because they are such `clean' systems to model; 
the inspiral waveform can be computed accurately to several post-Newtonian orders adequate for 
optimal signal extraction and parameter estimation. The typical strength of the source is:
\begin{equation}
h \sim 2.5 \times 10^{-23} \left(\frac{{\cal M}}{M_\odot}\right)^{5/3}  
\left(\frac{f}{{\rm 100~Hz}}\right)^{2/3} 
\left (\frac{r}{{\rm 100~ Mpc}} \right )^{-1} \,,
\end{equation} 
where ${\cal M}$ is the chirp mass equal to $(\mu M^{2/3})^{3/5}$, $\mu$ and $M$ being respectively 
the reduced and the total mass of the system, $r$ is the distance to the source - it is given 
at the scale of 100 Mpc because such events would be rare and therefore to obtain a reasonable 
event rate, a sufficient volume of the Universe needs to be covered - and $f$ is the instantaneous 
fiducial frequency of the source as the source evolves adiabatically during the inspiral stage. Since the phase 
of the waveform, apart from the amplitude, can be computed accurately by post-Newtonian methods, the optimal 
extraction technique of matched filtering is used. In the recent past, numerical relativity has been able to 
make a breakthrough by continuing the waveform to the merger phase and eventually connect it with the ringdown 
of the final  black hole. It is here that Chandrasekhar's contribution stands out because he pointed out that 
a black hole rings like a bell if it is subjected to a perturbation (Chandrasekhar \& Detweiler 1975). 
In the current context this occurs in the final stages of the merger when a black hole is formed. 
Quasi-normal modes were first discovered by Vishveshwara (1970) while examining the stability 
of the Schwarzschild black hole.  

Another important burst source of GW is supernovae.
It is difficult to reliably compute the waveforms for supernovae, because
complex physical processes are involved in the collapse and the resulting GW
emission. This limits the data analysis and optimal signal extraction.

Continuous wave sources pose one of the most computationally intensive
problems in GW data analysis 
(Schutz 1989; Brady et al. 1998; Cutler, Gholami \& Krishnan 2005).
A rapidly rotating asymmetrical neutron star is a source of continuous
gravitational waves.  There are some astrophysical systems known from
electromagnetic observations which might be promising sources of
continuous GWs.  Surveys for continuous GWs have so far not led to a
direct detection, but the searches have now become astrophysically
interesting.  We mention the result for the Crab pulsar in the next subsection.  
These searches for known systems are not computationally intensive since they target a
known sky position, frequency and spindown rate.  On the other hand,
blind all-sky and broad-band searches for previously unknown neutron
stars are a different matter altogether.  Long integration times,
typically of the order of a few months or years are needed to build up
sufficient signal power. The reason for this is that the signal is very weak and 
lies way below the detector noise level. We give a typical example:
\begin{equation}
h \sim 10^{-25} \left(\frac{I}{10^{45}{\rm gm.cm^2}}\right) 
\left(\frac{f}{1 {\rm kHz}}\right)^2 
\left(\frac{\epsilon}{10^{-5}}\right) 
\left (\frac{r}{10 {\rm kpc}} \right )^{-1} \,,
\end{equation}
where $I$ is the moment of inertia of the neutron star, $r$ the distance to the source, $f$ the 
GW frequency and $\epsilon$ is a measure of asymmetry of the neutron star. The asymmetry of a 
neutron star can occur in various ways such as crustal deformation, intense magnetic fields not 
aligned with the rotation axis or the Chandrasekhar-Friedman-Schutz instability 
(Chandrasekhar \& Esposito 1970; Friedman \& Schutz 1978). 
This instability is in fact driven by GW emission and consists of strong hydrodynamic waves in 
the star's surface layers. This phenomenon results in significant gravitational radiation.  
Earth's motion Doppler modulates the signal, and this Doppler modulation depends on the 
direction to the GW source. Thus, coherent extraction of the signal whose direction and 
frequency is unknown is an impossibly computationally expensive task.  The
parameter space is very large, and a blind survey requires extremely large computational resources.

To detect stochastic background one needs a network of detectors, ideally
say two detectors preferably identically oriented and close to one another. The 
stochastic background arises from a host of unresolved independent GW sources and can be 
characterised only in terms of its statistical properties. The strength of the source is 
given by the quantity $\Omega_{\rm GW} (f)$ which is defined as the energy-density of GW 
per unit logarithmic frequency interval divided by $\rho_{\rm critical}$, the energy density 
required to close the Universe. The typical strength of the Fourier component of the GW 
strain for the frequency bandwidth $\Delta f = f$ is:
\begin{equation}
{\tilde h}(f) \sim 10^{-26} \left(\frac{\Omega_{\rm GW}}{10^{-12}}\right) 
\left(\frac{f}{{\rm 10 Hz}}\right)^{-3/2} {\rm Hz}^{-1/2} \,,
\end{equation} 
The signal is extracted by cross-correlating the outputs. Two kinds of data-analysis 
methods have been proposed (i) a full-sky search - but this drastically limits the bandwidth 
(Allen \& Romano 1999), 
(ii) a radiometric search in which the sky is scanned pixel by pixel - since a small part 
of the sky is searched at a time, it allows for larger bandwidth, and more importantly 
includes the bandwidth in which the current detectors are most sensitive, thus potentially 
leading to a large SNR (Mitra et al. 2008). Moreover, with this method a detailed map of the 
sky is obtained.

Apart from these sources, there can be burst sources of GW from mergers or explosions or 
collapses which may or may not be seen electromagnetically but nevertheless deserve attention. 
In this case time-frequency methods are the appropriate methods which look for excess power 
in a given time-frequency box.  

In the section on data analysis, we will focus on two of the above mentioned and prominent GW 
sources, namely, the compact binaries and the continuous wave sources. Before moving on to 
the data analysis we would like to briefly describe the astrophysically interesting results 
so far obtained in GW astronomy.

\subsection{Astrophysical results from current GW data}
Even in this initial stage of the detectors, it is important to note that astrophysically 
interesting results have been obtained from the data so far taken with the LIGO detectors, 
more specifically, the data from the S5 run. The data have set astrophysically interesting 
upper limits on the GW emanating from astrophysical sources. We mention a few of the important 
results below.

The S5 data have constrained the cosmological GW background in which the upper limit falls 
below the previous upper limit set by nucleosynthesis (LIGO Science Collaboration and VIRGO Science 
Collaboration 2009). 
This result has excluded several string theory motivated big bang models. 

The GW data analysis from the S5 run shows that less than 4\% of the energy can be radiated away 
in GW from the Crab pulsar (Abbott et al. 2008a) 
This is because the spindown rate is $\sim 3.7 \times 10^{-10}$ Hz/sec, while 
no GW signal was observed even as low as $h \sim 2.7 \times 10^{-25}$.

Since no GW signal was detected from the GRB source 0702012, this implies that a compact binary progenitor with
masses in the range $1 M_\odot < m_1 < 3 M_\odot$ and $1 M_\odot < m_2 < 40 M_\odot$ located in M 31 is excluded as
a GW source with 99\% confidence. If the binary progenitor was not in M 31, then it rules out a binary star merger
progenitor upto a distance of 3.5 Mpc, with 90\% confidence 
and assuming random orientation (Abbott et al. 2008b). 
A search was performed from the LIGO S5 and the Virgo first science runs for the total mass
of the component stars ranging from 2 to 35 M$_\odot$. No GWs were identified. The 90 per cent confidence 
upper limit on the rate of coalescence of non-spinning binary neutron stars was 
estimated to be 8.7$\times$10$^{-3}$ yr$^{-1}$ L$_{10}^{-1}$,
where L$_{10}$ is 10$^{10}$ times the blue solar luminosity (Abadie et al. 2010).    

These are some of the salient astrophysical results which merely serve to indicate the 
revolutionary scientific impact that GW astronomy can bring to science.

\section{Data analysis of GW sources}
As can be seen from the foregoing, data analysis of interferometric data is a very important 
aspect in the quest for detection of gravitational waves. This is because the signal is weak 
and must be extracted from the noisy data - infact the noise, in general, strongly overwhelms the 
signal. Thus sophisticated statistical techniques and efficient algorithms based on statistical analysis 
are vital for extracting the signal from the noise. The data analysis technique of course 
depends on the nature of the signal. We would like to discuss a couple of sources and their data analysis 
in more detail. We first describe the matched filtering paradigm for the inspiraling binaries 
and then describe some current efforts in the so called `All sky all frequency search' for GW from 
periodic or continuous wave sources, which are based on group theoretic methods. This does not mean 
that the sources not discussed here are unimportant in any way, but the idea here is to give a 
flavour of the data analysis methods employed in GW detection. Here we have chosen two such data analysis problems. 

In this article we would like to emphasise the importance of the role of symmetries which play a 
crucial part in increasing the efficiency of an algorithm and in turn reducing the computational 
burden. The symmetries arise from the physical model of the GW source. The idea is to capture the 
symmetries in terms of group representation theory and then use the representation theory to 
develop efficient search algorithms.  

\subsection{Inspiraling/coalescing binaries} 
Here we will deal essentially with the inspiral waveform which is the first stage when the stars are 
relatively far apart, and the stage ends a little before the last stable orbit is reached. 
The last stable circular orbit for a test particle orbiting a Schwarzschild black hole of mass $M$ 
is at radial distance of $6MG/c^2$. Here we may take $M$ to be the total mass of the binary components, 
and then the inspiral stage is the one before the orbit shrinks to around $10 M$ or a little less. 
After the inspiral stage comes the merger stage, and the final stage is that when a single black hole 
is formed (in the case the masses are two black holes). Just before the final black hole is formed it oscillates, 
emitting quasi-normal mode radiation finally settling into a stable configuration of a stationary 
black hole. The merger waveform for black holes can now be computed from numerical relativity; 
there was a recent breakthrough in 2005 (Pretorius 2005), and this was followed by several groups 
actually implementing their numerical code (Campanelli et al. 2006; Baker et al. 2006). 
The inspiral waveform we will consider also holds for two neutron stars or a neutron star/black hole pair. Here we
will restrict ourselves to the binary inspiral and data analysis for it.  

\subsubsection{Matched filtering}
\begin{figure}
\centerline{\includegraphics[width=10cm]{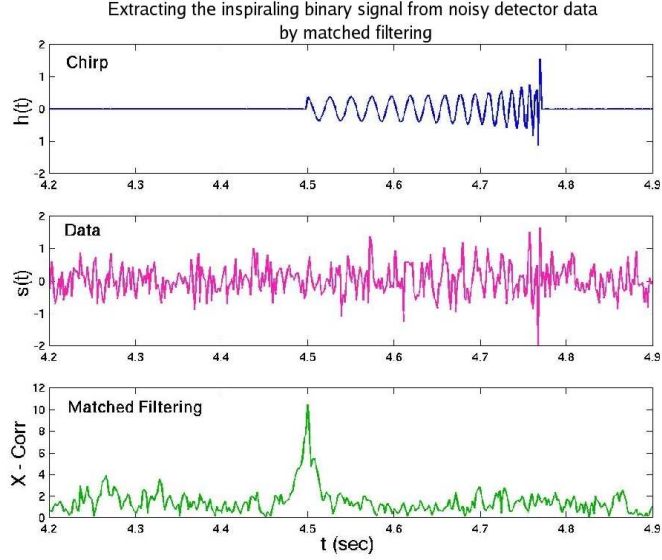}}
\caption{The top part of the figure shows the signal - the inspiral binary waveform usually called the chirp; the middle part shows the signal embedded in the detector noise while the bottom shows the plot of 
the output of the matched filter $c(\tau)$. By recognising the peak by thresholding, the signal can be detected.}
\label{MF}
\end{figure}

The appropriate technique to use, when one has the accurate knowledge of the waveform - 
especially of the phase - is matched filtering. First, it yields the maximum signal-to-noise (SNR) 
among all linear filters. Secondly, it is optimal in the Neyman-Pearson sense - in additive Gaussian 
noise, the matched filter statistic gives the maximum detection probability for a given false alarm 
rate. The matched filtering operation is defined as follows: if $x(t)$ is the data in the time domain, 
then the matched filter output $c(\tau)$ at the epoch $\tau$ is given by:
\begin{equation}
c(\tau) = \int x(t) q(t + \tau) dt \,,
\label{corr}
\end{equation}
where $q(t)$ is the matched filter. In stationary noise (the noise is independent of absolute time) 
$q$ has a particularly simple form and is conveniently described in the Fourier domain as:
\begin{equation}
{\tilde q} (f) = \frac{{\tilde h} (f)}{S_h (f)} \,,
\end{equation}
where $ h (t)$ is the expected signal in the detector and $S_h (f)$ is the power spectral density of 
the noise. An illustration of the matched filtering paradigm is given in Figure~\ref{MF}.

\subsubsection{Searching the parameter space: the spinless case}

In this section we will be considering only the point mass approximation which is valid for black holes and 
to a large extent for neutron stars if they do not deform. The problem would have been simple if there 
were a single signal waveform. But the signal depends on several parameters. Thus one is actually searching 
through a family of signals. The signal has the form:
\begin{equation}
h(t; \A, t_a, \phi_a, \tau_0, \tau_3) = \A a(t - t_a, \tau_0, \tau_3) \cos [\phi(t - t_a, \tau_0, \tau_3) + \phi_a] \,,
\label{sig}
\end{equation}
where $\A$ is the amplitude, $t_a$ is the time of arrival of the signal, $\tau_0$ and $\tau_3$ as defined below are
functions of the individual masses $m_1, m_2$ of the binary and $\phi_a$ the phase at arrival of the wave. The
signal described in equation~(\ref{sig}) is what is called the restricted post-Newtonian waveform in which the
amplitude is of the Newtonian waveform which is slowly varying with time, while the phase is given to as much as
accuracy is possible, that is upto the 3.5 post-Newtonian order which is deemed sufficient because it gives a phase
accuracy to better than a cycle for the stellar mass objects inspiraling in the bandwidth of the current or even
advanced detectors.  It is most important for the technique of matched filtering that the phase is known as
accurately as possible, because even half a cycle can put the signal waveform out of phase with the template
waveform which can lead to substantial decrease in the output of the matched filter. The amplitude $\A$ depends on
the chirp mass parameter $\mu M^{2/3}$ and on the fiducial frequency $f_a$ of the wave at the time of 
arrival $t_a$. Instead of the masses, it has been found useful to use the chirp times $\tau_0$ and $\tau_3$ as 
signal parameters - these parameters appear in a simple way in the Fourier transform of the signal, namely, they appear {\it linearly} in the phase of the Fourier transform in the stationary phase approximation. The final search algorithm becomes simple in terms of these parameters. They are related to $M$ and $\eta = \mu/M$ by the relations:
\begin{equation}
\tau_0 = \frac{5}{256\,\eta f_a} \left ( \pi M f_a\right )^{-5/3}, \,\,\,
\tau_3 = \frac{1}{8 \eta f_a} \left ( \pi M f_a \right )^{-2/3}.
\end{equation}
where $\tau_0$ is the Newtonian time of coalescence and the chirp time $\tau_3$ is related to 
the 1.5 PN order. As one can see from equation~(\ref{sig}), both the amplitude $a$ and phase $\phi$ depend 
on these parameters. We do not give the explicit forms of these functions here because they are unimportant to the
discussion here, but they can be found in the literature, eg. (Mohanty \& Dhurandhar 1996). 

We use the maximum likelihood approach, that is, the likelihood ratio must be maximised over the signal parameters, 
namely, $\{\A, t_a, \phi_a, \tau_0, \tau_3\}$. The maximum likelihood method shows that a matched filter is 
the simpler surrogate statistic than the likelihood ratio, and it is sufficient to maximise the output of 
the matched filter over the search parameters. The amplitude $\A$ is readily extracted from the signal by 
normalising the template waveform. The parameters $t_a$ and $\phi_a$ are searched for by using the symmetry 
of the signal family. The signal is {\it translationally invariant} in time, that is, a signal at another 
time of arrival is just obtained from translation. This symmetry can be exploited by using the 
Fast Fourier Transform (FFT), that is, writing equation~(\ref{corr}) in the Fourier domain:
\begin{equation}
c(t_a) = \int \frac{{\tilde x}^* (f) {\tilde h} (f)}{S_h (f)} e^{2 \pi i f t_a} df + {\rm {complex~conjugate}} \,,
\end{equation}    
where the integral in the Fourier domain is carried out essentially over the bandwidth of the detector and where
the signal cuts off at the upper frequency end. This is a family of integrals parametrised by all the signal
parameters, in particular, $t_a$. We have suppressed other parameters to avoid clutter, since we now focus on
$t_a$. But the $c(t_a)$ can be obtained for each $t_a$ by just using the FFT algorithm. This saves enormous
computational effort because now the number of operations reduces to order $N \log_2 N$ rather than $N^2$, where $N$ is the number of samples in the data segment. Typically, for a 500 sec. data train sampled at 2 kHz, $N \sim 10^6$, which implies a saving of computational cost of more than $10^4$! 
\par
\begin{figure}
\centerline{\includegraphics[width=8cm]{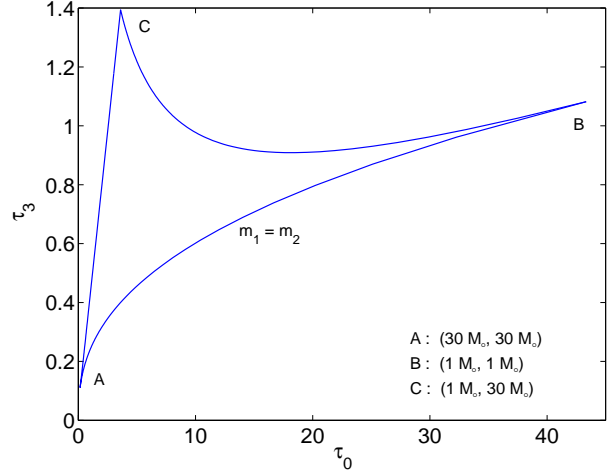}}
\caption{Parameter space in terms of the parameters $\tau_0, \tau_3$ for the mass range $1 M_\odot \leq m_1, m_2 \leq 30 M_\odot$ and $f_a = 40$ Hz.}
\label{bndry}
\end{figure}
\begin{figure}
\centerline{\includegraphics[width=8cm]{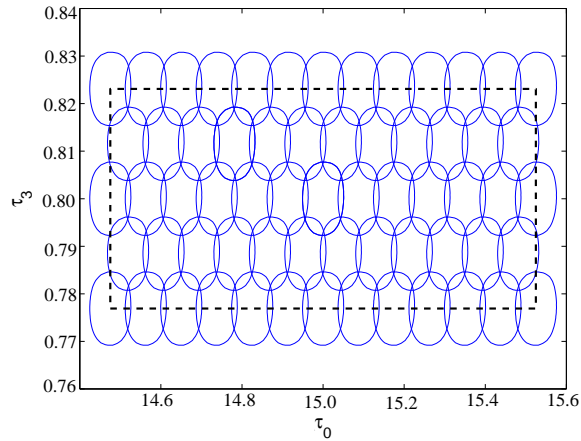}}
\caption{Hexagonal tiling of the parameter space.}
\label{hex}
\end{figure}

In the phase parameter $\phi_a$ also, there is a symmetry - changing $\phi_a$ to say $\phi_a' = \phi_a + \phi_0$ involves just adding a constant phase to the signal and the waveform still remains within the family. This is the so called $S^1$ symmetry. The search over phase can be carried out by using just two templates say for $\phi_a = 0$ and $\phi_a = \pi / 2$. If we call the correlations so obtained $c_0$ and $c_{\pi/2}$ respectively, where we have computed these correlations from the corresponding templates $h (f; \phi_a = 0)$ and $h (f; \phi_a = \pi/2)$, the $c(\phi_a)$ at arbitrary $\phi_a$ is then given by,
\begin{equation}
c(\phi_a) = c_0 \cos \phi_a + c_{\pi/2} \sin \phi_a \,,
\end{equation}
where we have suppressed other parameters to avoid clutter. Moreover, the maximisation of $c(\phi_a)$, the surrogate statistic, over $\phi_a$ can be done analytically. Thus,
\begin {equation}
\max_{\phi_a} c(\phi_a) = \left(c_0^2 + c_{\pi/2}^2 \right)^{1/2} \,.
\end{equation}
Thus the kinematical parameters $t_a$ and $\phi_a$ in the signal waveform are efficiently dealt with;    
the search over the masses, which are the dynamical parameters, now remains. There does not seem to be any 
efficient way, for example of using symmetries, to search over these parameters. Figure~(\ref{bndry}) shows 
the parameter space for $1 M_\odot \leq m_1, m_2 \leq 30 M_\odot$ in the parameters $\tau_0$ and $\tau_3$. 
Since the waveform is symmetric in $m_1, m_2$, one needs to only search the space $m_1 \leq m_2$. This gives  
roughly a triangular shape to the search region of the parameter space which is topologically equivalent to 
the triangle in $(m_1, m_2)$ space.
One now spans the parameter space densely with a bank of templates. The templates are arranged so that the 
maximum mismatch between a signal and a template never exceeds a small fixed amount. 
The usual number is taken to be 3\% which corresponds to a maximum loss of 10\% in the event rate of the signals. 
With this criterion, in the parameters $\tau_0, \tau_3$ the templates are approximately uniformly spaced. 
The idea is to {\it tile} the parameter space so that (i) there are no `holes' and (ii) there is minimal overlap, 
so that the number of templates is reduced to a minimum which then in turn reduces the computational burden. 
The best such scheme happens to be hexagonal tiling as shown in Figure~\ref{hex}.
One does this template placement elegantly by defining a metric 
(Balasubramanian, Sathyaprakash \& Dhurandhar 1996; Owen 1996) over the parameter space. Then one 
finds that the metric coefficients are nearly constant when the parameters $\tau_0, \tau_3$ are used; 
the parameters in fact play the role of coordinates on a signal manifold, and the above statement can 
be reworded as saying that $\tau_0, \tau_3$ are like Cartesian coordinates, while $m_1, m_2$ are curvilinear. 
For this level of mismatch the number of templates required is $\sim 10^4$ for the noise PSD of the initial LIGO. If the signal is cut-off at a little less than 1 kHz, so that the sampling rate is 2 kHz, then a simple computation shows that the online search for these signals is little more than 3 GFlops.      

One now takes the maximum of the matched filter output over all the parameters and compares this maximum with a
threshold. The threshold is set by the noise statistics and the false alarm rate that one is prepared to tolerate.
Clearly, the false alarm rate must be much less than the expected event rate. If the noise is Gaussian, the $c$
maximised over $\phi_a$ has a Rayleigh distribution in the absence of the signal. Assuming a false alarm rate of 1 per year, gives a false alarm probability of $\sim 10^{-14}$ for one year observation period, which in turn sets the threshold at 8.2 (this is in units of the standard deviation of the Gaussian noise). Detection is announced if the $c$ maximised over the parameters exceeds the threshold. However, in order to achieve good detection probability one must have $c$ well over the threshold - thus $c > 8.9$ gives a detection probability better than about 95\%. 

The foregoing describes the general matched filtering paradigm. It clearly holds for a larger mass range or a
larger bandwidth. If one lowers the lower limit of the band to 10 Hz as will be the case for advanced detectors and
increases the mass range to begin from say $0.2 M_\odot$, the online search requirement increases by a hundred
times.  Also if one includes spins, the waveform now must depend on 6 more parameters, namely, the spin vectors
$\vec{S}_1, \vec{S}_2$, and the computational burden increases roughly by three orders of magnitude. 

In order to deal with the rather large computational cost, hierarchical schemes have been proposed 
(Mohanty \& Dhurandhar 1996; Sengupta, Dhurandhar \& Lazzarini 2003)
which can reduce this cost. The idea is to look for triggers with a low threshold with a coarse bank of templates,
and then follow only the trigger events by a fine search and then use the high threshold as in the regular search
described above. This method can reduce the cost considerably and theoretical factors of few tens in reducing the
cost have been obtained in stationary Gaussian noise. These reduction factors of course will be significantly less
in the presence of real detector noise. Recently Cannon et al. (2010) have shown that singular value decomposition can be used to reduce the computational cost. They have shown that for neutron star/neutron star inspiral where the parameter space is smaller, the computational gain can be almost an order of magnitude. 

\subsubsection{Merger and ringdown}
For a black hole merger, one must solve Einstein's equations with complex initial and boundary conditions. 
Due to the nonlinearity of Einstein's equations, the problem is highly complex, and work had been 
going on for few decades before
Pretorius (2005) made a breakthrough. This was followed by other successful numerical solutions 
(Campanelli et al. 2006; Baker et al. 2006). Clearly, the first such solutions corresponded to nonrotating 
black holes which now are not too difficult to obtain. The rather surprising fact was that the merger phase is 
a smooth continuation of the inspiral phase, contrary to what had been expected. There are results also for spinning 
black holes. Work is in progress to obtain numerical relativistic solutions spanning the entire parameter space for different mass ratios and spins. There are also interesting effects such as `kicks' in the general case; the final black hole has a residual linear momentum.

The final ring-down phase consists of a superposition of quasi-normal modes (QNM) with their amplitude depending on the details of the perturbation. But each QNM is uniquely given by the black hole mass and the angular momentum parameter. The `no hair' theorem for black holes in general relativity states 
that a black hole is completely characterised by its mass and angular momentum. The above mentioned property of QNMs is a consequence of this theorem. Thus observing QNMs would unambiguously show that the source is a black hole and also confirm the no hair theorem of general relativity.   

The important point for data analysis is that the inspiral waveform can be continued into the merger phase and to the ring-down phase of the final black hole to obtain a single stitched waveform, thus yielding a higher SNR. The mass range can now go upto $100 M_\odot$ and the distance by about a factor of 2 which would then correspondingly increase the event rate by about an order of magnitude.
These searches are currently being performed by the Ligo Science Collaboration.  

\subsection{The all sky, all frequency search for GW from rotating neutron stars}

 We will consider the simple model of an isolated rapidly rotating neutron star and ignore spindown. We will show
here, how the group theory and other algebraic methods can be used to elegantly formulate the problem by exploiting
the symmetries in the physical model. In this endeavour, we will make use of the \emph{stepping around the sky
method} - a method proposed by Schutz more than twenty years ago (Schutz 1989), which gives an apt framework for 
this approach. There have been a host of methods proposed, notably the Hough transform, the stack and slide, and 
resampling methods (Schutz \& Papa 1999; Pletsch \& Allen 2009; Patel et al. 2010) which reduce the computational cost over the 
straight-forward search over 
the sky direction, frequency, and spindown parameters. Although these methods significantly reduce the computational burden, it is not reduced to the point where the search can be performed with the current computer resources available in reasonable time. Therefore, it becomes necessary to explore novel approaches which address this problem.

Consider a \emph{barycentric frame} in which the isolated neutron star is at rest or moving with uniform velocity. 
Ignoring spindowns the signal in this frame is assumed to be a pure sinusoid - monochromatic of constant frequency 
say $f$. The detector however, takes part in an accelerated motion - in general a superposition of simple harmonic 
motions of varying amplitudes and phases. Thus the signal in the detector is not a pure sinusoid but is modulated 
by Doppler effects - the Doppler correction depending on the direction to the source, relative to the motion of 
the detector. Since the detector moves in a complicated way relative to the barycentre, a complex Doppler profile 
is generated which depends on the direction to the source. If the direction to the source and the frequency are 
unknown, the Doppler profile is unknown and then one must face the problem of scanning over all directions 
in the sky and also over frequency. From astrophysical considerations usually the maximum frequency $f_{\rm max}$ 
is taken to be 1 kHz. The \emph{stepping} method gives a direct way for obtaining the Fourier transform 
in the barycentric frame of the demodulated signals connecting two different directions say $\n$ and $\n'$.    

The signal expected is so weak that one typically needs to integrate the signal for several months or a year 
before one can build up significant SNR. So if the observation time needed is $T \sim 10^7$ sec or more 
and if the maximum frequency $f_{\rm max}$ to be scanned is taken about a kHz, then the number of samples in a data
train are $N \sim 2 f_{\rm max} T \sim 10^{10}$. Since the detector orbits the Sun in this time, the `aperture' of
the `telescope' is the diameter $D$ of the Earth's orbit; $D \sim 3 \times 10^8$ km while the minimum GW wavelength
is $\lambda_{\rm GW} \sim 300$ km corresponding to a frequency of 1 kHz. Thus the resolution is $\Delta \theta =
\lambda_{\rm GW}/D \sim 10^{-6}$ radian, a fraction of an arc second - the Fourier transform of such a signal
spreads into a million Fourier bins, and consequently the signal is lost in the noise of the detector. One
therefore needs to demodulate the signal first and then take its Fourier transform in order to collect the signal
power in a single frequency bin. This means one needs to scan or demodulate over $N_{\rm patches} \simeq 4 \pi/
{\Delta \theta}^2 \sim 10^{13}$ directions or patches in the sky. So even this naive calculation gives the number
of operations for the search to be $N_{\rm ops} \sim 3 N_{\rm patches} N \log_2 N \sim 10^{25}$ if one were to
perform the FFT of the data train after demodulating in each direction. A machine with a speed of few teraflops
would need several thousand years to perform the analysis! Moreover, this estimate excludes overheads, and ignores
spindown parameters. Including these would increase the cost of the search by several orders of magnitude. Thus
the search is highly computationally expensive, and novel and original ideas should be explored, if this search 
has to be brought within the capabilities of current resources or those envisaged in the near future. 
The approach outlined here is based on group theory and is one such attempt towards finding a  solution to this problem.

Moreover, there exists also the possibility of using this approach in tandem with the previous approaches which 
have been aimed at reducing the computational cost. It is envisaged that a judicious combination of several 
approaches may go towards alleviating the computational burden.
 
\subsubsection{The formulation of the problem}
Let the motion of the detector be described in general by $\R (t)$ in the barycentric frame $(X, Y, Z)$; 
for circular motion $\R (t) = R (\cos \Om t~ \X + \sin \Om t~ \Y )$ - we take the detector motion in the $(X, Y)$ plane - where $\X$ and $\Y$ are unit vectors along the $X$ and $Y$ axes respectively, and $\Om$ is the angular velocity of the detector in the barycentric frame.  We will treat $\R (t)$ generally for now until later when we specialise to circular motion. The key defining equation which describes the transformation between barycentric time $t$ and detector time $t'$ is:
\begin{equation}
s' (t') = s (t) \,.
\label{defn}
\end{equation}
The detector time coordinate $t'$, which is in fact a retarded or advanced time, is given by $t' = t - \R (t) \cdot \n / c$ and is related to the barycentric time coordinate $t$. From our assumptions, the signal in the barycentric frame can be taken to be monochromatic. So after demodulation a Fourier transform is all that is necessary to extract the signal from the detector noise. It is in fact the matched filter!

We write the Doppler modulation in an abstract form in terms of an operator. The signal at the detector coming from the direction $\n$ is related to the signal in the barycentric frame by the equation:
\begin{equation}
{\bf s}' (\n) = M (\n) ~ {\bf s} \,,
\label{fwd}
\end{equation}
where $M (\n)$ is the modulation operator which is defined via equation~(\ref{defn}). This operator has explicit
representations in the time as well as in the  Fourier domain (Dhurandhar \& Krishnan 2011). Note that in the 
all sky, all frequency search we do not know $\n$. Therefore we need to scan over the directions. A trial demodulation is performed for some general direction $\n'$  given by $\n' = (\sin \theta' \cos \phi', \sin \theta' \sin \phi', \cos \theta')$, which is not necessarily $\n$. Thus we try the direction $\n'$ and have a trial demodulated signal, 
\begin{equation}
{\bf s}_{\rm trial} (\n'; \n) = M^{-1} (\n')~ {\bf s}' (\n) \,.
\label{rvrs}
\end{equation}
If $\n' \neq \n$, then the demodulation is incorrect and we must try again with a different $\n'$ until we get to $\n$ or atleast get close enough. If $\n' \simeq \n$, we must observe a peak in Fourier domain. Using these formulae we can now \emph{step directly} to a direction $\n'$ as follows:
\begin{equation}
{\bf s}_{\rm trial} (\n'; \n) = Q (\n', \n) ~{\bf s} \,,
\label{bary}
\end{equation}
where the \emph{stepping} operator is defined by: 
\begin{equation}
Q (\n', \n) = M^{-1} (\n') ~M (\n) \,.
\label{ker}
\end{equation}
This was the approach suggested by Schutz, now expressed in our formulation, so that one may directly \emph{step} from the direction $\n$ to the direction $\n'$ in the space of demodulated waveforms. This formulation was expected to enhance the efficiency of the search for example by using the sparseness of the matrices. The approach here builds upon this formulation. Apart from the sparseness of matrices, the idea is to use symmetries in the problem for stepping efficiently in the sky. The symmetry is made manifest via the language of group theory. 

In order to get a group structure and go beyond the method advocated by Schutz, it is necessary to expand the scope
of the direction vectors $\n$ to the full three dimensional Euclidean space ${\cal R}^3$. It is clear that this is
required because even a step in the sky namely, $\n' - \n$ will not be of unit length. Thus it is necessary as also
convenient to `unwrap' the space of directions, which is a projective space, to its universal covering space ${\cal
R}^3$. We then define the operators $M ({\bf a})$, where ${\bf a}$ is an arbitrary vector in ${\cal R}^3$, and
${\bf a}$ is used in the `retarded time' instead of $\n$. We can then show that these operators now form a group,
atleast approximately, well within the physical requirements (Dhurandhar \& Krishnan 2011). It is important to note
that these operators $M$ act on functions, namely signals, and map them to other signals - the signals are Doppler
shifted. Such groups are called transformation groups in the literature (Vilenkin 1988). 

\subsubsection{Circular motion of the detector}

  For concreteness, we give an example of circular motion of the detector. This is a very simplified case because
in reality the detector partakes of a complicated superposition of simple harmonic motions which have complex set of phases. This simple case is taken to see how the group theory helps. The group now is reduced to Euclidean group in 2 dimensions, usually denoted by $E(2)$. We consider the motion as above and consider the situation when the motion consists of exactly one orbit, i.e. $0 \leq t \leq T$ and $\Om T = 2 \pi$. Then in the Fourier space, where $n = f/T$ and $n' = f'/T$, we look at the action of $M ({\bf a})$ on the complete orthonormal basis of the Hilbert space of square integrable functions over $[0, T]$, namely, the set of functions $e^{2 \pi i n t/T}$. The natural scalar product on this Hilbert space for the two functions $g_1$ and $g_2$ is defined by:
\begin{equation}
(g_1, g_2) = \frac{1}{T} \int_0^T g_1(t)~g_2^*(t)~dt \,.
\end{equation} 
In this basis, the matrix representation for $M$ (we have chosen ${\bf a} = \n$ a unit vector) is readily given:
\bea
M (\n; n', n) &=& (M (\n)~e^{2 \pi i n \frac{t'}{T}}, ~e^{2 \pi i n' \frac{t'}{T}}) \, \no \\
               &=& (e^{2 \pi i n \frac{t}{T}}, ~e^{2 \pi i n' \frac{t'}{T}}) \, \no \\
               &=& \frac{1}{T} \int_0^T dt' ~e^{2 \pi i n \frac{t}{T} - 2 \pi i n' \frac{t'}{T}} \,. 
\eea
where we have used the definition $M (\n) s(t') = s(t)$. 
An explicit expression for $M(\n; n', n)$ can be obtained for the direction $\n = (\sin \theta \cos \phi, \sin \theta \sin \phi, \cos \theta)$. Writing $\psi = \Om t'$ and $\beta = R \Om / c$ we obtain:
\bea
M (\n; n', n) &=& \frac{1}{2 \pi} \int_{0}^{2 \pi} d \psi~ e^{i(n - n') \psi + i n \beta \sin \theta \cos (\psi  - \phi)} \, \no \\
&\equiv& e^{i \chi(n - n')}~J_{n - n'} \left (n \beta \sin \theta \right) \,,
\eea
where $\chi = \phi + \pi/2$ is the translated azimuthal angle. 
\par
From the form of $M (\n; n', n)$ it is evident that when applied to the data vector $x_n$, the search in $\chi$ can be performed by a fast Fourier transform; the stepping in the azimuthal parameter is done in an efficient way. If there are $B$ samples of the $\chi$ parameter, then the search over $\chi$ for a given $\theta$ and frequency $n'$ can be performed in order of $B \log_2 B$ number of operations. It may be further possible to reduce the number of operations by similar methods, but this example underlines the role of symmetry and the group theory in developing efficient data processing algorithms. 

\section{Concluding Remarks}

The era of gravitational wave astronomy has arrived. The initial detectors have not only reached their promised
sensitivities but have surpassed them. The advanced detectors will start operating in few years time and the era of
gravitational wave astronomy would then have truly begun. From the astrophysical knowledge that we possess as of
now, one should expect a fair rate of gravitational wave events that one should be able to observe. An important
recent development has been LIGO-Australia where LIGO is planning to build one of its detectors in Australia with
partial funding from Australia. A detector far  away and out of the plane of other detectors in US and Europe would
greatly benefit the search of gravitational waves. In this India is also thinking of chipping in, so that India
also has a stake in this exciting world project. Already, India has a 20 year old legacy in gravitational wave data
analysis at IUCAA, Pune and wave form modelling at RRI, Bengaluru, and recently a three metre prototype detector at T.I.F.R., Mumbai has been funded. Apart from the groundbased detectors, there is also the prospect of the space-based ESA-NASA detector LISA which will bring in important astrophysical information at low frequencies complementing the ground-based detectors. The future looks bright for GW astronomy.     



\label{lastpage}
\end{document}